\documentclass[multphys,vecphys]{svmult}

\usepackage{makeidx}         % allows index generation
\usepackage{graphicx}        % standard LaTeX graphics tool
\usepackage{multicol}        % used for the two-column index
\usepackage[bottom]{footmisc}% places footnotes at page bottom
\usepackage{amssymb}

\begin{document}

\title*{Extragalactic Relativistic Jets and Nuclear Regions in Galaxies}
\titlerunning{Relativistic Jets and Nuclear Regions in Galaxies} 
\author{Andrei Lobanov
	\and
	Anton Zensus
	}
\authorrunning{Lobanov \& Zensus} 
\institute{Max-Planck-Institut f\"ur Radioastronomie, Auf dem H\"ugel 69,
53121 Bonn, Germany.
\texttt{alobanov@mpifr-bonn.mpg.de}, \texttt{azensus@mpifr-bonn.mpg.de}}
\maketitle

\begin{abstract}
Past years have brought an increasingly wider recognition of the ubiquity
of relativistic outflows (jets) in galactic nuclei, which has turned
jets into an effective tool for investigating the physics of nuclear
regions in galaxies. A brief summary is
given here of recent results from studies of jets and nuclear regions
in several active galaxies with prominent outflows.
\end{abstract}

\section{Introduction}
\label{lobanov2:sec1}

Substantial progress achieved during the past decade in studies of
active galactic nuclei (see~\cite{lobanov2006} for a review of recent
results) has brought an increasingly wider recognition of the ubiquity
of relativistic outflows (jets) in galactic
nuclei~\cite{falcke2001,zensus1997} turning them into an effective
probe of nuclear regions in galaxies~\cite{lobanov2005}. Emission
properties, dynamics, and evolution of an extragalactic jet are
intimately connected to the characteristics of the supermassive black
hole, accretion disk and broad-line region in the nucleus of the host
galaxy~\cite{lobanov2006}.  The jet continuum emission is dominated by
non-thermal synchrotron and inverse-Compton
radiation~\cite{unwin1997}. The synchrotron mechanism plays a more
prominent role in the radio domain, and the properties of the emitting
material can be assessed using the turnover point in the synchrotron
spectrum~\cite{lobanov1998b}, synchrotron
self-absorption~\cite{lobanov1998a}, and free-free absorption in the
plasma~\cite{kadler2004,walker2000}.

High-resolution radio observations access directly the regions where
the jets are formed~\cite{junor1999}, and trace their evolution and
interaction with the nuclear environment~\cite{mundell2003}. Evolution
of compact radio emission from several hundreds of extragalactic
relativistic jets is now systematically studied with dedicated
monitoring programs and large surveys using very long baseline
interferometry (such as the 15\,GHz VLBA\footnote{Very Long Baseline
Array of National Radio Astronomy Observatory, USA}
survey~\cite{kellermann2004} and MOJAVE~\cite{lister2005}).  These
studies, combined with optical and X-ray studies, yield arguably the
most detailed picture of the galactic nuclei. Presented below is a
brief summary of recent results from studies outlining the relation
between jets, supermassive black holes, accretion disks and broad-line
regions in prominent active galactic nuclei (AGN).

\begin{table}[t]
\caption{Characteristic scales in the nuclear regions in active galaxies}
\label{lobanov2:tb1}
\begin{center}
\begin{tabular}{rccccc}\hline\hline
   & $l$ & $l_8$ & $\theta_\mathrm{Gpc}$ & $\tau_c$ & $\tau_\mathrm{orb}$ \\ 
  & [$R_\mathrm{g}$] & [pc] & [mas]& [yr] & [yr] \\ \hline
Event horizon:           &1--2          &$10^{-5}$           &$5\times 10^{-6}$    &0.0001     & 0.001 \\
Ergosphere:              &1--2          &$10^{-5}$           &$5\times 10^{-6}$    &0.0001     & 0.001 \\
Accretion disk:          &10$^1$--10$^3$&$10^{-4}$--$10^{-2}$&$0.005$              &0.001--0.1 & 0.2--15 \\
Corona:                  &10$^2$--10$^3$&$10^{-3}$--$10^{-2}$&$5\times 10^{-3}$    &0.01--0.1& 0.5--15 \\
Broad line region:       &10$^2$--10$^5$&$10^{-3}$--1        &$0.05$               &0.01--10   & 0.5--15000 \\
Molecular torus:         &$>$10$^5$     &$>$1                &$>$$0.5$             &$>$10      & $>$15000 \\
Narrow line region:      &$>$10$^6$     &$>$10               &$>$5                 &$>$100     & $>$500000 \\
Jet formation:           &$>$10$^2$     &$>$$10^{-3}$        &$>$$5\times 10^{-4}$ &$>$0.01    & $>$0.5 \\
Jet visible in the radio:&$>$10$^3$     &$>$$10^{-2}$        &$>$$0.005$           &$>$0.1     & $>$15 \\ \hline
\end{tabular}
\end{center}
{\bf Column designation:} $l$ -- dimensionless scale in units of the
gravitational radius, $G\,M/c^2$; $l_8$ -- corresponding linear scale,
for a black hole with a mass of $5\times 10^8\,$M$_{\odot}$;
$\theta_\mathrm{Gpc}$ -- corresponding largest angular scale at 1\,Gpc
distance; $\tau_c$ -- rest frame light crossing time;
$\tau_\mathrm{orb}$ -- rest frame orbital period, for a circular
Keplerian orbit. Table is reproduced from~\cite{lobanov2006}
\end{table}

\section{Anatomy of jets}

Jets in active galaxies are formed in the immediate vicinity of the
central black hole, and they interact with every major
constituent of AGN (see Table~\ref{lobanov2:tb1}). The jets carry away
a fraction of the angular momentum and energy stored in the accretion
flow~\cite{hujeirat2003} and in the rotation of the central black
hole~\cite{koide2002,komissarov2005,semenov2004}. At distances of $\sim
10^3\,R_\mathrm{g}$, the jets become visible in the radio regime,
which makes high-resolution VLBI observations a tool of choice for
probing directly the physical conditions in AGN on such small
scales~\cite{junor1999,krichbaum2004}. Recent studies indicate that at
$10^3$--$10^5\,R_\mathrm{g}$ ($\lesssim 1$\,pc) the jets are likely to
be dominated by pure electromagnetic processes such as Poynting
flux~\cite{sikora2005} or have both MHD and electrodynamic
components~\cite{meier2003}. The magnetic field is believed to play an
important role in accelerating and collimating extragalactic jets on
parsec scales~\cite{vlahakis2004}. Three distinct regions
with different physical mechanisms dominating the observed properties
of the jets can be identified: 1)~ultracompact jets where collimation
and acceleration of the flow occurs; 2)~parsec-scale flows dominated
by relativistic shocks and 3)~large-scale jets where plasma
instabilities are dominant.

\subsection{Ultracompact jets}

Ultracompact jets observed on sub-parsec scales typically show
strongly variable but weakly polarized emission (possibly due to
limited resolution of the observations). In many cases, the emission
is optically thick, indicating that opacity effects may play a
prominent role~\cite{lobanov1998a}.  Ultracompact outflows are
probably dominated by electromagnetic
processes~\cite{meier2003,sikora2005}, and they become visible in the
radio regime (identified as compact ``cores'' of jets ) at the point
where the jet becomes optically thin for radio
emission~\cite{lobanov1998a,lobanov1999}. The ultracompact jets do not
appear to have strong shocks~\cite{lobanov1998b}, and their evolution
and variability can be explained by smooth changes in particle density
of the flowing plasma, associated with the nuclear flares in the
central engine~\cite{lobanov1999}. Compelling evidence exists for
acceleration~\cite{bach2005} and collimation~\cite{junor1999,krichbaum2004} of the
flows on these scales.

\subsection{Parsec-scale flows: shocks and instabilities}

Parsec-scale outflows are characterized by pronounced curvature of
trajectories of superluminal
components~\cite{kellermann2004,lobanov1999}, rapid changes of
velocity and flux density and predominantly transverse magnetic
field~\cite{jorstad2005}. Relativistic shocks are prominent on these
scales, which is manifested by strong polarization~\cite{ros2000} and rapid
evolution of the turnover frequency of synchrotron
emission~\cite{lobanov1997}.  Mapping the turnover frequency
distribution provides also a sensitive diagnostic of plasma
instabilities in relativistic flows~\cite{lobanov1998b}.  Complex
evolution of shocked regions is revealed in
observations~\cite{gomez2001,jorstad2005,lobanov1999} and numerical
simulations~\cite{agudo2001} of parsec-scale outflows. However, the
shocks are shown to dissipate rapidly~\cite{lobanov1999}, and shock
dominated regions are not likely to extend beyond $\sim
100$\,pc. Starting from these scales, instabilities (most importantly,
Kelvin-Helmholtz instability) determine at large the observed
structure and dynamics of extragalactic
jets~\cite{lobanov2001,lobanov2003,perucho2006}.  Non-linear evolution
of the instability~\cite{perucho2004a,perucho2004b} and stratification
of the flow~\cite{perucho2005} are important for reproducing the
observed properties of jets.  Similarly to stellar jets, rotation of
the flow is expected to be important for extragalactic
jets~\cite{fendt1997}, but observational evidence remains very limited
on this issue.

\section{Jets and nuclear regions in AGN}

A number of recent studies have used jets to probe physical conditions
in the central regions of AGN.  Opacity and absorption in the nuclear
regions of AGN have been probed effectively using the non-thermal
continuum emission as a background source~\cite{lobanov2005}. The
free-free absorption studies indicate the presence of dense, ionized
circumnuclear material with $T_\mathrm{e} \approx 10^4$\,K distributed
within a fraction of parsec from the central
nucleus~\cite{lobanov1998a,walker2000}. 

Absorption due to several atomic and molecular species (most notably
due to H\i, CO, OH, and HCO$^+$) has been detected in a number of
extragalactic objects. OH absorption has been used to probe the
conditions in warm neutral gas~\cite{goikoechea2004,kloeckner2005},
and CO and H\i\ absorption were used to study the
molecular tori~\cite{conway1999,pedlar2004} at a linear resolution
often smaller than a parsec~\cite{mundell2003}. These observations
have revealed the presence of neutral gas in a molecular torus in
NGC\,4151 and in a rotating outflow
surrounding the relativistic jet in 1946$+$708~\cite{peck2001}.

Connection between accretion disks and relativistic
outflows~\cite{hujeirat2003} has been explored, using correlations
between variability of X-ray emission produced in the inner regions of
accretion disks and ejections of relativistic plasma into the
flow~\cite{marscher2002}. The jets can also play a role in the
generation of broad emission lines in AGN. The beamed continuum
emission from relativistic jet plasma can illuminate atomic material
moving in a sub-relativistic outflow from the nucleus, producing broad
line emission in a conically shaped region located at a significant
distance above the accretion disk~\cite{arshakian2006}. Magnetically
confined outflows can also contain information about the dynamic
evolution of the central engine, for instance that of a binary black
hole system~\cite{lobanov2005b}. This approach can be used for
explaining, within a single framework, the observed optical
variability and kinematics and flux density changes of superluminal
features embedded in radio jets.

\section{Conclusion}

Extragalactic jets are an excellent laboratory for studying physics of
relativistic outflows and probing conditions in the central regions of
active galaxies. Recent studies of extragalactic jets show that they
are formed in the immediate vicinity of central black holes in
galaxies and carry away a substantial fraction of the angular momentum
and energy stored in the accretion flow and rotation of the black
hole. The jets are most likely collimated and accelerated by
electromagnetic fields. Relativistic shocks are present in the flows
on small scales, but dissipate on scales of $\lesssim 100$\,pc. Plasma
instabilities dominate the flows on larger scales.  Convincing
observational evidence exists connecting ejections of material into
the flow with instabilities in the inner accretion disks. In
radio-loud objects, continuum emission from the jets may also drive
broad emission lines generated in sub-relativistic outflows
surrounding the jets.  Magnetically confined outflows may preserve
information about the dynamics state of the central region, allowing
detailed investigations of jet precession and binary black hole
evolution to be made. This makes studies of extragalactic jets a
powerful tool for addressing the general questions of physics and
evolution of nuclear activity in galaxies.

%%%%%%%%%%%%%%%%%%%%%%%%%%%%%%%%%%%%%%%%%%%%%%%%%%%%%%%%%%%%%%%%%%%%%%  

%%%%%%%%%%%%%%%%%%%%%%%%%%%%%%%%%%%%%%%%%%%%%%%%%%%%%%%%%%%%%%%%%%%%%%

%\printindex
\end{document}